\documentstyle[prd,preprint,aps,epsfig]{revtex}
\begin{document}

\draft
\def\be{\begin{equation}}
\def\ee{\end{equation}}
\def\bea{\begin{eqnarray}}
\def\eea{\end{eqnarray}}
\def\nn{\nonumber}
\def\ep{\epsilon}
\def\c{\cite}
\def\m{\mu}
\def\ga{\gamma}
\def\lan{\langle}
\def\ran{\rangle}
\def\Ga{\Gamma}
\def\thet{\theta}
\def\la{\lambda}
\def\Lam{\Lambda}
\def\ka{\chi}
\def\si{\sigma}
\def\al{\alpha}
\def\pa{\partial}
\def\de{\delta}
\def\De{\Delta}
\def\Dex{\Delta_{,x}}
\def\Dey{\Delta_{,y}}
\def\Dev{\Delta^{-1}}
\def\Ome{\Omega}
\def\Om2{\Omega^{2}}
\def\ov{\over}
\def\gmn{g_{\mu\nu}}
\def\gmnv{g^{\mu\nu}}

\def\rsr{{r_{s}\over r}}
\def\rrs{{r\over r_{s}}}   
\def\rs2r{{r_{s}\over 2r}}
\def\l2r2{{l^{2}\over r^{2}}}
\def\rsa{{r_{s}\over a}}
\def\rsb{{r_{s}\over b}}
\def\rsro{{r_{s}\over r_{o}}}
\def\rss{r_{s}}
\def\a2{{l^{2}\over a^{2}}}
\def\b2{{l^{2}\over b^{2}}}
\def\op{\oplus}
\def\sn{\stackrel{\circ}{n}}
\def\c{\cite}

%\twocolumn[\hsize\textwidth\columnwidth
%\hsize\csname @twocolumnfalse\endcsname

\title
{Rotation intrinsic spin coupling---the parallelism description}

\author{C. M. Zhang } 
\address{
Department of Physics, National Tsing Hua University, Hsinchu 300,       Taiwan\\
% Tel: 886-3-5742545
% Fax: 886-3-5723052
zhangcm99@yahoo.com,
zhangcm@phys.nthu.edu.tw}

\author{A. Beesham}
\address{Department of Mathematical Sciences,
 University of Zululand,   
 Kwa-Dlangezwa 3886,  South Africa}

%\date{\today}

\maketitle

\begin{abstract}

For the Dirac particle in the rotational system, 
 the rotation induced inertia effect is analogously treated as 
the modification of  the ``spin connection" on the Dirac equation 
in the flat spacetime, which
is determined by the equivalent tetrad. 
From the point of view of parallelism description of spacetime, 
the obtained torsion axial-vector is just the rotational 
angular velocity, which is included in the ``spin connection". 
Furthermore the axial-vector spin coupling induced spin 
precession 
is just the rotation-spin(1/2) interaction predicted by 
Mashhoon. Our derivation treatment is straightforward and simplified 
in the geometrical meaning and physical conception, however the 
obtained conclusions are consistent with that of 
the other previous work.

\end{abstract}
\pacs{04.25.Nx, 04.80.Cc}
\vskip1pc]

%\newpage

\section{Introduction}

Recently the spin-rotation-gravity coupling has 
been paid much attention and appeared in the work
of many authors who have been mainly
interested in the study of wave equations
in accelerated systems and gravitational fields 
\c{mas88,hn,chap,huang,tiom96,ryd97,ryd99,ryd20}.
Indeed, the coupling under consideration here directly
involves wave effects that pertain to the
physical foundations of general relativity.
It follows that similar rotation-spin coupling effects are expected
in a rotating frame of reference\c{mas88,mas20}.  

The observational consequences of rotation-spin coupling 
for neutron interferometry in a
rotating frame of reference have been explored 
in connection with the assumptions that underlie
the physical interpretation of wave equations 
in an arbitrary frame of reference \c{masexp,sted85}.  In
general, the rotation-spin phase shift is smaller 
than the Sagnac shift \c{sted97} by roughly the
ratio of the wavelength to the dimension of the interferometer.

A proper theoretical treatment of the inertial properties
of a Dirac particle is due to Hehl
and Ni \c{hn}.
This treatment has been extended in several important directions
by a number of investigators \c{hn,huang,ryd99,ryd20}.
The significance of rotation-spin coupling for atomic
physics has been pointed out by Silverman \c{silv}.

Moreover, the astrophysical consequences of the 
helicity flip of massive neutrinos as a consequence 
of rotation-spin coupling have been
investigated by Papini et al. \c{papi}.
Furthermore, the influence of the rotation-spin coupling 
 on the magnetic inclination evolution of pulsars 
has also been investigated\c{zcm92,zcm98}.

The direct evidence for the coupling of intrinsic spin to
the rotation of the Earth has recently
become available\c{masexp}.  In fact, according to the
natural extension of general relativity
under consideration here,
every spin-$1\over 2$ particle in the laboratory has an additional
interaction Hamiltonian.   As
measured by the observer, however,
such intrinsic spin must ``precess''
in a sense opposite to the sense of rotation of
the Earth. The Hamiltonian associated with such motion 
would be of the form \c{mas20}
\be
\de H = - \mbox{\boldmath$\Omega$} 
\cdot \mbox{\boldmath$\sigma$}\;,
\label{ham}
\ee
where $\mbox{\boldmath$\Omega$}$ is the frequency of rotation
of the laboratory frame.
The existence of such a Hamiltonian would show 
that intrinsic spin has rotational inertia.

In quantum mechanics, mass and spin characterize the
irreducible unitary representations of the
inhomogeneous Lorentz group.
The inertial properties of mass are well known
in classical mechanics through various translational and rotational
acceleration effects.
  It is therefore interesting to consider the inertial
properties of spin \c{mas88,masexp}.

 The aim of the present paper is to discuss the rotation-spin  
 effect in straightforward way, i.e., discussing the inertia 
 effect on the Dirac equation by means of the parallelism 
 description of spacetime, so the article is organized as follows, 
in Section II,
we introduce the  teleparallel equivalent description of  general
relativity (GR), 
and in  Section III, 
we discuss the Dirac equation in GR and in the framework of 
parallelism description. 
 In  Section IV, 
 we extend our discuss to the rotational  system, where 
 the torsion axial-vector induced 
 spin precession interaction are studied.
 The conclusions and further expectation of the teleparallel equivalent
 description of the axisymmetrical  spacetime    
 will appear  in Section V. 
 We use the unit in which 
 the  speed of light is set equal to unit: $c = 1$.

\section{the teleparallel equivalent of general relativity}

The teleparallel equivalent of general relativity (PGR) 
has been pursued by a number of authors ~\cite{per1,mal,hay79}, 
 where the spacetime  is characterized by the
 torsion tensor and the 
 vanishing  curvature, the relevant spacetime is the
Weitzenb\"ock
spacetime~\cite{hay79}, which is a special case of the Riemann-Cartan
spacetime with the constructed  metric-affine theory of
gravitation. 
As is well known, at least in the absence of spinor fields,
the teleparallel gravity is equivalent to general relativity. 
We will use the greek alphabet ($\mu$, $\nu$, $\rho$,~$\cdots=1,2,3,4$)
to denote tensor indices, that is, indices related to spacetime. The latin alphabet
($a$, $b$, $c$,~$\cdots=1,2,3,4$) will be used to denote local Lorentz (or tangent space)
indices. Of course, being of the same kind, tensor and local Lorentz indices can be
changed into each other with the use of the tetrad $h^{a} {}_{\mu}$, which satisfy
\be
e^{a}{}_{\mu} \; e_{a}{}^{\nu} = \delta_{\mu}{}^{\nu} \quad
; \quad e^{a}{}_{\mu} \; e_{b}{}^{\mu} = \delta^{a}{}_{b} \; .
\label{orto}
\ee
A nontrivial tetrad field can be used to define the linear Cartan connection 
\be
\Gamma^{\sigma}{}_{\mu \nu} = e_a{}^\sigma \partial_\nu e^a{}_\mu \;
,
\label{car}
\ee 
with respect to which the tetrad is parallel:  
\be {\nabla}_\nu \; e^{a}{}_{\mu}
\equiv \partial_\nu e^{a}{}_{\mu} - \Gamma^{\rho}{}_{\mu \nu}
\, e^{a}{}_{\rho} = 0 \; . 
\label{weitz}
\ee 
The Cartan connection can be decomposed according to 
\be
{\Gamma}^{\sigma}{}_{\mu \nu} = {\stackrel{\circ}{\Gamma}}{}^{\sigma}{}_{\mu
\nu} + {K}^{\sigma}{}_{\mu \nu} \; ,
\label{rel} 
\ee
where
\be
{\stackrel{\circ}{\Gamma}}{}^{\sigma}{}_{\mu \nu} = \frac{1}{2}
g^{\sigma \rho} \left[ \partial_{\mu} g_{\rho \nu} + \partial_{\nu}
g_{\rho \mu} - \partial_{\rho} g_{\mu \nu} \right]
\label{lci}
\ee
is the Levi--Civita connection of the metric 
\be
g_{\mu \nu} = \eta_{a b} \; e^a{}_\mu \; e^b{}_\nu \; ,
\label{gmn}
\ee
and 
\be {K}^{\sigma}{}_{\mu \nu} = \frac{1}{2}
\left[ T_{\mu}{}^{\sigma}{}_{\nu} + T_{\nu}{}^{\sigma}{}_{\mu}
- T^{\sigma}{}_{\mu \nu} \right]
\label{conto}
\ee 
is the contorsion tensor, with 
\be
T^\sigma{}_{\mu \nu} =
\Gamma^ {\sigma}{}_{\mu \nu} - \Gamma^{\sigma}{}_{\nu \mu} \;  \label{tor} 
\ee
the torsion of the Cartan connection~\c{hay79}. The irreducible
torsion vectors, i.e., the torsion vector and 
the torsion axial-vector, can then be
constructed as~\c{hay79} 
\be
V_{\mu} =  T^{\nu}{}_{\nu \mu}
\ee
\be
A^{\m} = {1\over 6}\ep^{\m\nu\rho\si}T_{\nu\rho\si}
\ee

{}The nontrivial tetrad field induces both, a riemannian and a teleparallel
structures in spacetime. The first is related to the Levi--Civita
connection, a connection presenting curvature, but no torsion. The second
is related to the Cartan connection, a connection presenting torsion, but
no curvature. It is important to remark that both connections are defined
on the very same spacetime, a spacetime endowed with both a riemannian and
a teleparallel structures.

\section{Dirac equation in the curved spacetime}

The gravitational effects on the spin incorporated into Dirac
equation  through the ``spin
connection'' $\Gamma_{\mu}$ appearing in the Dirac equation
in curved spacetime \cite{dirac},
which is constructed by means of the variation of the covariant Lagrangian
of the spinor field as,
 
\be
\left[ \gamma^{a} e^{\mu}_{a} (\partial_{\mu} + \Gamma_{\mu})
        + m\right] \psi = 0.
\label{dirac2}
\ee
The explicit expression for $\Gamma_{\mu}$ can be written in terms of
the Dirac matrices and tetrads(see also ~\cite{ful96})
\be
\Gamma_{\mu} \equiv {1 \over 8}[\gamma^b, \gamma^c] e_{b}{}^{\nu} 
        e_{c \nu;\mu}.
\label{ga1}
\ee
We must first simplify the Dirac matrix product in the spin
connection term. It can be shown that
\be
\gamma^a [\gamma^b, \gamma^c] = 2 \eta^{ab} \gamma^c -
        2 \eta^{ac} \gamma^b - 2i \epsilon^{dabc} \gamma_5
        \gamma_d,  \label{gammas}
\ee
where $\eta^{ab}$ is the metric in flat space and
$\epsilon^{abcd}$ is the (flat space) totally antisymmetric tensor,
with $\epsilon^{0123}= +1$. With Eq.(\ref{gammas}), the
contribution from the spin connection is arranged as   
\be
\Ga_{\mu} \equiv
 {1\over 2}V_{\mu}  - {3i \over 4}A_{\mu}\ga_{5},
\label{ga2}
\ee
which means that Eq.(\ref{ga1}) and Eq.(\ref{ga2}) are equivalent but 
just the different mathematical form\c{hay79}. 
 Alternatively completed in the parallelism description of the 
Weitzenbock spacetime (see Ref.\c{hay79}), 
 Dirac equation can be obtained by the variation method, 
 which is constructed by means of the variation of 
 the covariant Lagrangian
 of the spinor field. The  Dirac Langrange density
$L_{D}$ is regularly given by
   
\be
L_{D} = {1 \over 2}e^{\mu}_{k}[{\psi}\gamma^{k}\nabla_{\mu}\bar{\psi}
- \nabla_{\mu}\bar{\psi}\gamma^{k}\psi ] - m \bar{\psi} \psi.
\ee
By taking variation with respect to $\bar{\psi}$,  the Dirac equation in
Weitzenb\"ock spacetime is given as described in Eq.(\ref{dirac2}) with 
the spin connection in  Eq.(\ref{ga2}). It is interesting to note 
that the 
torsion  axial-vector represents the deviation of the axial symmetry
from the spherical symmetry \cite{nit80}.
In Weitzenb\"ock spacetime, as well as the general version of torsion 
gravity, it has been shown by many
authors~\c{hay79,nit80} that the spin
precession of a Dirac particle  is intimately
related to the torsion axial-vector,
\be
\frac{d{\bf S}}{dt} = - {3\ov2} \mbox{{\boldmath $A$}} \times {\bf S}
\label{precession1}
\ee
where ${\bf S}$ is the spin vector of a Dirac particle, 
and $\mbox{{\boldmath $A$}}$ is the spacelike part 
of the torsion axial-vector.
Therefore, the corresponding extra Hamiltonian is of the form, 

\be
\de H = - {3\ov2} \mbox{{\boldmath $A$}}\cdot \mbox{\boldmath$\sigma$}
\label{ham2}
\ee

\section{The rotation-spin effect in the parallelism description}

Now we discuss the  Dirac equation in the 
rotational coordinate system, and imagine a rotating disk with the 
angular velocity $\Ome$ in the experimental laboratory system, 
at where the Dirac particle locates, we set the rotation axis 
in z-direction. 
In the  rotational  coordinate system (t,x,y,z),
 the tetrad components can be obtained from 
the following line element (c.f. \c{hn,chap})
\bea
ds^{2}&=& g_{\mu\nu}dx^{\mu}dx^{\nu} \nn \\
      &=& \De dt^{2} + 2\Ome y dtdx  - 2\Ome x dtdy \nn \\
      &-& (dx^{2} + dy^{2} + dz^{2}  )\;, 
\label{dsr}
\eea
where $\De = 1 - \Ome^{2} r^{2}$ and $r^{2} = x^{2}+ y^{2}$. 
In the matrix form, the metric and its inverse are written as, 
\be
\gmn {} = \pmatrix{
\De    & \Ome y  & -\Ome  x & 0 \cr
\Ome y  & -1      & 0       & 0 \cr
-\Ome  x &0      & -1       & 0 \cr
0      &0       & 0       &-1} ,
\ee 

\be
\gmnv {}  =  \pmatrix{
1    & \Ome y  & -\Ome  x & 0 \cr
\Ome y  & -1+ \Om2 y^2      & -\Om2 xy       & 0 \cr
-\Ome  x & -\Om2 xy       & -1 + \Om2 x^2       & 0 \cr
0      &0       & 0       & -1} ,
\ee 
\be
g=det|\gmn|= -1\;.
\ee

The tetrad can be obtained with the subscript 
$\mu$ denoting the column index (c.f. \c{hn}),  

\be
e^{a}{}_{\mu} = \pmatrix{
1    & 0 & 0 & 0 \cr
-\Ome y & 1 & 0 & 0 \cr
\Ome  x & 0 & 1 & 0 \cr
0      & 0 & 0 & 1} ,
\label{te1}
\ee

with the inverse 
$e_{a}{}^{\mu} = \gmnv e^{b}{}_{\nu} \eta_{ab}$

\be
e_{a}{}^{\mu} = \pmatrix{
1 & \Ome y  & -\Ome x  & 0 \cr
0 & 1 &0  &0\cr
0 &0 &1  &0 \cr
0 &0 &0  &1} .
\label{te2}
\ee
We can inspect that 
Eqs.(\ref{te1}) and (\ref{te2}) satisfy the conditions in 
Eqs.(\ref{orto}) and (\ref{gmn}).     
Or equivalently, the tetrad can be expressed by the 
dual basis of the differential one-form \c{hn} through  
choosing a coframe of the rotational coordinate  system, 

\bea
  \vartheta ^{\hat{0}}& =& \,  d\,t\;, \\
\quad\vartheta ^{\hat{1}} &=&\, dx - \Ome y dt \;,\\
 \quad\vartheta ^{\hat{2}} &=&\,  dy + \Ome x dt\;,\\
\quad\vartheta ^{\hat{3}}&=& \, dz \;,
\eea
with the obtained metric as 
\begin{equation}
ds^2=\eta_{ab}\,\vartheta^{a}\otimes\vartheta^{b}
\label{metric}
\end{equation}
is in agreement with Eq.(\ref{dsr}) and that in Ref.\c{hn}. 
 {}From Eqs.(\ref{te1}) and (\ref{te2}), we can now construct the
Cartan connection, whose nonvanishing components are:
\be
\Ga^{2}{}_{01} =   \Ome,  \;\; \;\;
\Ga^{1}{}_{02} = - \Ome, \;\;
\ee
 The corresponding nonvanishing torsion components are:

\be
T^{2}{}_{01} =  \Ome ,  \;\;
T^{1}{}_{02} =  - \Ome , 
\ee
The torsion vector and the axial torsion-vector are consequently
\be
V_{\m}  = 0,   \;\; \m=0,1,2,3\;, 
\ee

\be
A_{3} = {2\ov 3}\Ome \; ,   \;\; A_{k} = 0, k=0,1,2\;.
\ee

As shown, $ A_{1}  = A_{2} = 0$ is  on account of the Z-axis
symmetry which results in the canceling of the x and y 
components, and then generally we can write 
$\mbox{{\boldmath $A$}} = {2\ov3}{\bf \Ome}$ and the corresponding 
additive Hamiltonian induced by the axial-vector spin coupling 
in Eq.(\ref{ham2})  
$\de H=-{\bf \Ome}\cdot \mbox{\boldmath$\sigma$}$, which is expected 
in Eq.(\ref{ham}) by Mashhoon.  
 From the spacetime geometry view, 
the torsion axial-vector represents the deviation from the spherical
symmetry~\c{nit80}, i.e., which will  disappear in the spherical case 
(Schwarzschild spacetime for instance) and occurs in the axisymmetry 
case (Kerr spacetime for instance). 
Therefore the torsion axial-vector 
corresponds to a inertia field with respect to Dirac particle, 
which is now explictly expressed by Eq.(\ref{precession1}) that 
\be
\frac{d{\bf S}}{dt} = -  \mbox{{\boldmath $\Ome$}} \times {\bf S}\;,
\label{precession2}
\ee
which is same as that expected by Mashhoon (c.f. Ref.\c{mas20}).

\section{Discussions and conclusions}
The inertia effect on the Dirac
particle is studied in this work in the framework of the 
parallelism description of spacetime. In particular,
these results are valid for a neutron 
and a mass neutrino. Therefore, the rotation-spin coupling, predicted 
by Mashhoon for a neutron wave, has been derived in an alternative 
way. 
We recovered the rotation-spin effect in a straightforward derivation, 
by means of the parallelism description of spacetime, and 
the rotation-spin effect  can be clearly expressed by the 
spin precession effect of the  
irreducible torsion  axial-vector, which is constructed by the 
Cartan connection directly. The ``noninertia force" on Dirac particle  
can be  
preferably treated  as a rotation induced 
torsion of spacetime. 
Furthermore the constant axial-vector (angular velocity) means that 
 the ``noninertia force" is universally same in 
any spacetime position. 
However the  geometrical and physical 
meaning of the latter is simply and clearly shown. 
In the parallelism description of 
spacetime, the basic element of spacetime 
is a tetrad, and the metric is a by-product 
 and constructed by the tetrad\c{hay79}, however this 
fact is in priority to connect the Dirac equation because 
the ``spin connection" of Dirac equation is described by 
the tetrad directly but not by the  metric. The  
``spin connection"  can be decomposited into two irreducible 
torsion vectors, i.e., torsion vector and torsion axial-vector, 
and the latter represents the axisymmetry. 
The verification and consistency of our derivation in the 
the parallelism treatment of the inertia effect on Dirac particle 
leads us to believe that this description would be equivalently 
extended into the gravitomagnetic effect on Dirac particle 
\c{mas20}, 
where Kerr spacetime induced spin coupling will be examined.  
     
\section*{appendix}
%{Dirac Matrix  in parallel rotation spacetime}

Dirac matrix in curved spacetime can be given by the standard Dirac-Pauli
matrix in the Lorentz coordinates

\be
{\bf \al}\equiv \pmatrix {0 & {\bf \si}  \cr {\bf \si} &0 }
,\,\,\,\beta\equiv \pmatrix{0 & 1  \cr -1 &0 }
\ee
where ${\bf \si} = (\si_{x},\; \si_{y}, \;\si_{z})$ is Pauli matrix.
\be
{\si_{x}}\equiv\pmatrix{0 & 1  \cr 1 &0 }
,\,\,\,{\si_{y}}\equiv\pmatrix{0 & -i  \cr i &0 }
,\,\,\, {\si_{z}}\equiv\pmatrix{1 & 0  \cr 0 &-1 }
\ee
   
\be 
{\bf \ga}_{i}\equiv  \beta{\bf \al } = \pmatrix{0 & {\bf \si}_{i}  \cr
-{\bf \si}_{i} & 0 }
,\,\,\, \ga_{o} \equiv \beta \equiv \pmatrix{0 & 1  \cr -1 &0 }
\ee

The Dirac matrix in the 
 rotational  coordinates  can be expressed by the 
  standard Dirac-Pauli matrix in local Lorentz coordinates 
(the subscripts  are shown with parentheses),  
\bea
\ga_{o}&=& \ga_{(o)} - \Ome[ y\ga_{(1)} -  x\ga_{(2)}] \\ \nn
\ga_{1}&=&  \ga_{(1)} \\ \nn
\ga_{2}&=&  \ga_{(2)}\\ \nn
\ga_{3}&=&  \ga_{(3)}
\eea

\section*{Acknowledgments}

The author would like to thank J.G. Pereira  for discussion, and this work was 
supported by NSC of Taiwan and NRF of South Africa.

\end{document}